\documentclass[
showkeys,preprintnumbers,amsmath,amssymb,secnumarabic]{revtex4}
\usepackage{graphicx}
\usepackage{dcolumn}
\usepackage{bm}
\usepackage{epsfig}
\usepackage{float}
\usepackage{amsmath,mathtools}
\usepackage{setspace}
\usepackage[title]{appendix}

\usepackage{tensind}
\tensordelimiter{?}

\usepackage{setspace}
\usepackage{color}
\usepackage{dsfont}
\usepackage{mathrsfs}

\usepackage{stackengine}
\stackMath

\usepackage{graphicx}
\usepackage{tikz}
\usetikzlibrary{shapes.geometric, arrows}

\begin{document}

\renewcommand{\thesection}{\arabic{section}}
\renewcommand{\thesubsection}{\thesection.\arabic{subsection}}
\renewcommand{\thesubsubsection}{\thesubsection.\arabic{subsubsection}}

\newcommand{\bs}{\boldsymbol}
\newcommand\eqn[1]{\begin{eqnarray} #1 \end{eqnarray}}
\newcommand\vect[1]{\boldsymbol{#1}}
\newcommand\mat[1]{\mathsf{#1}}
\newcommand\trans{^\mathsf{T}}

\newcommand\braket[3]{\left<#1\,\right|#2\left|\,#3\right>}
\newcommand\ket[1]{\left|\,#1\right>}
\newcommand\bra[1]{\left<#1\,\right|}
\newcommand\braketsc[2]{\left<#1\,|\,#2\right>}
\newcommand\eg{\emph{e.g. }}
\newcommand\ie{\emph{i.e. }}
\newcommand\nn{\mathrm{N}_2}
\newcommand\hh{\mathrm{H}_2}
\newcommand\methane{\mathrm{CH}_4}
\newcommand\ris{\{\vect{r}_i\}}
\newcommand\rris{\{\vect{R}_i\}}
\newcommand\dris{d\vect{r}_i^N}
\newcommand\x{\vect{x}}
\newcommand\xis{\{\vect{x}_i\}}
\newcommand\xxis{\{\vect{X}_i\}}
\newcommand\dxis{d\vect{x}_i^N}

\tikzstyle{rect}  =[draw, rectangle, fill=white!20, text width=5.9 cm, text centered,   minimum height=2em,   minimum width=5.em]
\tikzstyle{rectx}=[draw, rectangle, fill=white!20, draw=white!20,      text centered,   minimum height=2em,   minimum width=0.2em]
\tikzstyle{rect2}=[draw, rectangle, fill=white!20,  text width= 6.cm,  text centered,   minimum height=2em,   minimum width=5 em]
\tikzstyle{line}=[draw, -latex']

\hyphenation{geo-metrization}

\title{Gravitational quantum dynamics: a geometrical perspective}
\author{Ivano Tavernelli} 
\email{ita@zurich.ibm.com}
 \affiliation{IBM Research -- Zurich, 8803 R\"uschlikon, Switzerland}
\date{\today}

\begin{abstract}
We present a gravitational quantum dynamics theory that combines quantum field theory for particle dynamics  in space-time with classical Einstein's general relativity in a non-Riemannian Finsler space.
This approach is based on the geometrization of quantum mechanics proposed in ref.~\cite{tavernelli_geom} and combines quantum and gravitational effects into a global curvature of the Finsler space 
induced by the quantum potential associated to the matter quantum fields.
In order to make this theory compatible with general relativity, the quantum effects are described in the framework of quantum field theory, 
where a covariant definition of `simultaneity' for many-body systems is introduced through the definition of a suited foliation of space-time.
As in Einstein's gravitation theory, the particle dynamics is finally described by means of a geodesic equation in a curved space-time manifold.
\end{abstract}


\maketitle

\section{Introduction}
The quest for a theory that reconciles quantum theory with general relativity~\cite{Einstein1915} (GR) has attracted the interest of many researchers since the early sixties of the past century.  
However, no single theory has yet emerged as the leading one and, on the contrary, we have witnessed a proliferation of different approaches that lead to the development of new fields of research in mathematics and theoretical physics.
A thorough summary of the different approaches is certainly beyond the purpose of this study and therefore we orient the interested readers towards the more specialized literature.

Following the `classification' proposed in~\cite{Rovelli2004}, the attempts to combine quantum theory with gravitation led to the development of several distinct, but also interconnected, theoretical models. 
The most promising among these candidates for quantum gravity is probably string theory. 
From a fundamental prospective string theory is developed from the quantum field theory of one-dimensional objects, the strings, which take the place of the original point particles. 
One of the vibrational states of the strings corresponds to the graviton, the quantum mechanical particle that carries the gravitational force~\cite{Becker2007}. %
Alternatively, other consistent theories of quantum gravity are obtained from the quantization, with different flavours, of the gravitational field and corresponding metric tensor, or of the space-time itself.
The most successful among these theories is probably loop quantum gravity~\cite{Rovelli1998,Carlip2001}. 
In this case, space-time is quantized producing a granularity of space (quantum of space), which defines a minimum scale distance through which matter can travel, known as Planck length.
At this scale, space is conceived as a `tissue' of finite loops also known as spin networks, whose dynamics produces spin foams. 
Mathematically, these objects correspond to a generalization of the Feynman diagrammatic perturbation theory where instead of a graph, a higher dimentional 2-complex topological space is used~\cite{Reisenberger1997,Engle2008}.
In addition to these two main research lines, there have been several other directions proposed, which include: Euclidean quantum gravity~\cite{DeWitt1967} based on Wick-rotation of the Minkowski space; twistor theory~\cite{Penrose1967}, which maps Minkowski space to a new geometrical object in a complex coordinate space known as twistor space, and noncommutative quantum field theories that play an important role in M-theory~\cite{Connes1998}.
Despite the beauty of the formalisms, none of these alternative theories has developed into a firm physical model of quantum gravity but they are mainly confined into the realm of mathematical hypotheses with limited experimental evidences.  

To these well established theories for quantum gravity we also need to add a series of alternative approaches that aim at deriving a relativistic covariant formulation of quantum theory in the framework of Bohmian mechanics~\cite{Bohm1952,Bohm1952a,Bohm_Hiley} and its field theory extension~\cite{hatfieldbook}. 
The first attempt to derive a fully consistent covariant version of Bohmian dynamics is due to Bohm \textit{et al.}~\cite{Bohm1987} and was later followed by the extensions developed by Holland~\cite{Holland1993,hollandbook}.
They showed that a consistent covariant formulation of Bohmian field theory is possible for both scalar and spinor fields leading to a new framework for a possible unification of quantum theory with gravitation.
Further investigations aiming at combining Bohmian quantum effects with the gravitation potential have also been proposed~\cite{Nikolic_2005,Castro_2006,Perelman2019}.
More recently, D\"urr and coworkers~\cite{Duerr2009}  made an additional step in this direction showing how a relativistic space-time Bohmian theory of many-body dynamics can be achieved using a privileged foliation of space-time, the same that we will use in our approach.
However, also in this case the purpose was to demonstrate the compatibility of Bohmian dynamics with GR and not the one of deriving a consistent theory for gravitational quantum dynamics (GQD).

In this work, we will take a different path and propose a theory of GQD~\cite{Quantum_Gravity_Def} where the quantum fields produce a further curvature (in addition to the one induced by the energy-matter tensor) of an extended space-time through the action of the quantum potential, which then guides the time evolution of point particles along geodesic paths.
This new theory for GQD is based on three fundamental pillars: the many-body field theory of relativistic particles, the Bohmian theory of quantum potential applied to field theories, and the extension of space-time (pseudo-) Riemann geometry  to Finsler geometry~\cite{Rund59,Bucataru2007,PhysRevD.75.064015}. 
The main reasons for this last point are at least twofold: 
First, due to the explicit dependence of the quantum potential from the time variable, the Lagrangian formulation of quantum mechanics as a point particle theory (in the spirit of de Broglie and Bohmian dynamics) requires an extended formalism that has in the Finsler geometry its most natural geometrical realization. 
Second, while in non-relativistic quantum mechanics a theory dependent on the time parameter is permissible, its extension to the relativistic case (Lorentz invariant) requires a parameter-independent formalism. 
This condition leads to a class of homogeneous Lagrangians with a differential-geometric representation as a Finsler space and length measure
\begin{equation}
L[\gamma]=\int d \tau \, F(\gamma, \dot \gamma)
\end{equation}
for a curve $\gamma: \, \tau \mapsto \gamma(\tau)$ in the base manifold, where $F(x, y)$ is the Finsler function homogeneous of degree one in the second variable $\dot \gamma=\partial \gamma/\partial \tau$. 
Finsler geometry includes Riemann and Minkowskian geometries as a special cases~\cite{Rund_1966}.
All choices made in the course of the derivation of this formalism are motivated by the following physical constraints: \textit{i}) recover the correct non-relativistic quantum mechanical limit, which was demonstrated in a previous publication~\cite{tavernelli_geom}, and \textit{ii}) ensure that the classical (non-quantum) limit of the theory coincides with Einstein's GR.

In  section~\ref{sec_QM}, we will review a trajectory-based non-relativistic quantum dynamics approach formulated in a Finsler space.  
Section~\ref{sec_Dirac} deals with the many-body (Dirac) theory in the relativistic space-time manifold putting particular emphasis on the definition of a covariant concept of `simultaneity' for GR~\cite{Duerr1999,Duerr2014,Misnerbook}. This will allow us to define unambiguously the concept of a relativistic invariant many-body quantum potential in a foliation of space-time.
Section~\ref{sec_Finsler} will be devoted to the description of Finsler geometry in the multi-time extended phase space, where the quantum potential enters in the definition of the metric tensor and the corresponding non-linear Cartan connection.
Finally, in section~\ref{sec_action} we will derive Einstein's field theory in the extended Finsler space and describe its connection to the original classical theory formulated by Einstein in 1915~\cite{Einstein1915}. A summary of the main steps is given in Fig.~1.

\begin{figure} [h]

\begin{center}
\begin{tikzpicture}[node distance = 1. cm, scale=0.9, transform shape]

\node [rect]   
(step1) {Non-relativistic quantum Lagrangian in Finsler space, $L_q$ [Eq.~\eqref{prop2b}]};

\node [rect, below of=step1, node distance= 1.5cm]  
(step11) {Covariant many-particle Dirac field equation through foliation, $\mathcal{F}$, of space-time};

\node [rectx, right of=step11, node distance=3.75 cm] (stepx1) {};
\node [rectx, right of=stepx1, node distance=3 cm] (step21) {};

\node [rect,  below of=step11, node distance= 1.5cm]  (step12) {Lorentz-invariant quantum potential on particle $i$, $Q(\bar{x}_N, x_i)$ [Eq.~\eqref{eq_quantum_field_potential}]};
\node [rectx, below of=stepx1, node distance=1.5 cm] (stepx2) {};
\node [rectx, below of=step21, node distance= 1.5 cm] (step22) {};

\node [rect,  below of=step12, node distance= 1.5cm]  (step13) {Covariant Finsler functions, 
$F_0$ [Eq.~\eqref{Finsler_0i}] and 
$F_{\mathscr{I}}$  [Eq.~\eqref{cq_action_inter}] 
with metrics
$\tilde{g}^0$ [Eq.~\eqref{metric_vac}]
and
$\tilde{g}$ [Eq.~\eqref{cq_gF}]};
\node [rectx, below of=stepx2, node distance=1.5 cm] (stepx3) {};
\node [rect, below of=step22, node distance= 1.5 cm] (step23) {General relativity: Covariant action, $\mathcal{A}_g$ [Eq.~\eqref{Einsten_action}], in Riemann space with metric $g$};

\node [rect2, below of=stepx3, node distance= 1.7 cm] (stepx4) {Combined action $\mathcal{A}^F_{g}(\tilde{g})$ [Eq.~\eqref{action}]};

\node [rect2, below of=stepx4, node distance= 1.5 cm] (stepx5) {Field equations in Finsler space with combined quantum and \\ gravitational effects [Eqs.~\eqref{cq_einstein1}-\eqref{cq_einstein2}]};

\path [line] (step1) -- (step11);
\path [line] (step11) -- (step12);
\path [line] (step12) -- (step13);
\path [line] (step23) -- (step13);
\path [line] (step23) -- (stepx4);
\path [line] (step13) -- (stepx4);
\path [line] (stepx4) -- (stepx5);

\end{tikzpicture}
\caption{Diagram summarizing the logical flow of this work.}
\end{center}
\label{figure_diagram}
\end{figure}
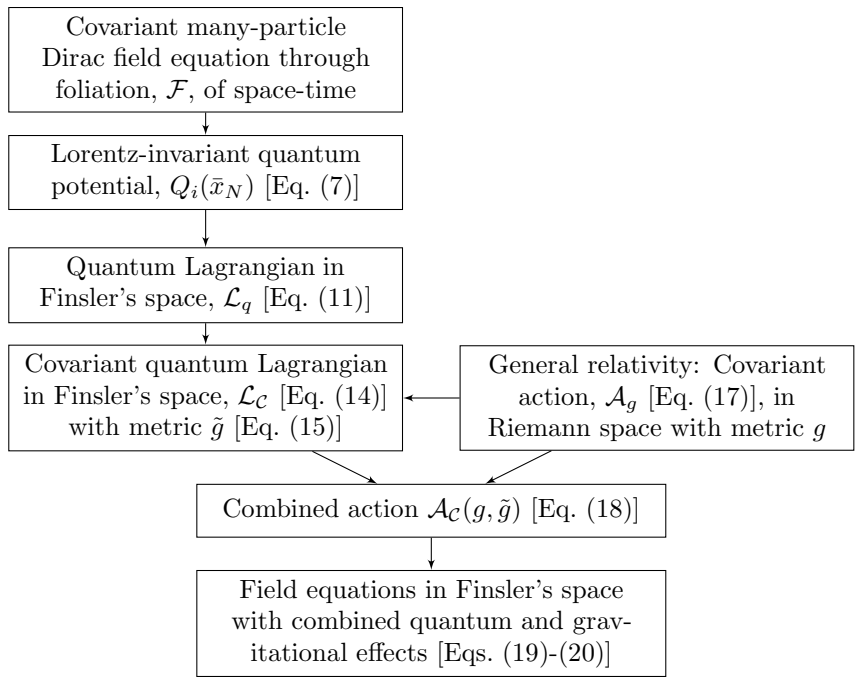

\section{Preamble: from non-relativistic Bohmian trajectories to geodesics} \label{sec_QM}
Recently, a new geometrical description of quantum dynamics based on Finsler geometry was introduced, which defines the metric tensor as a function of the positions and velocities in the cotangent bundle~\cite{tavernelli_geom}.
Within this framework, particles evolve along geodesic paths in a curved phase space manifold where the curvature is induced by the action of the quantum potential. 
The theory is fully deterministic and does not imply any form of probabilistic interpretation as the wavefunction nature of quantum theory is absorbed into the geometry of the space. 

At this point, it is convenient 
 to move to the multi-time framework already introduced in~\cite{tavernelli_geom}. 
The dimension of the multi-time phase space is therefore $2\times 4N$.
The generalization of the non-relativistic geodesic dynamics to the multi-time relativistic case (see next section) requires the formulation of the equation of motion of each individual particle, separately. 
In the following, we will use the variable $x_{\alpha}=(x_1,\dots,x_N)$ with $x_i=(x_i^0, x_i^1, x_i^2, x_i^3)$ for the positions and $y_{\alpha}=(y_1,\dots,y_N)$ with $y_i=(\dot x_i^0, \dot  x_i^1, \dot  x_i^2, \dot  x_i^3)$ for the corresponding velocities,  with $\dot x_i^a=\partial x^a_i/ \partial s$ ($a \in \{0,1,2,3\}$) where $s$ is the global time variable (proper time)~\cite{tavernelli_geom}.
For a trajectory in phase space we associate the geodesic curve $s \mapsto x(s)=\zeta(s)$~\cite{Rund59,tavernelli_geom} with particle components $(\zeta^0(s),\dots,\zeta^{(4N-1)}(s))$ and velocities $s \mapsto y(s)=\dot\zeta(s)$  (using Einstein's summation convention and $\alpha, \beta, \gamma, \delta=0,\dots,4N-1$), 
\begin{equation}
\ddot{\zeta}^\alpha + ?N^{\alpha}_{}\beta?(\zeta(s),\dot\zeta(s)) \dot\zeta^\beta= 
- {g}^{\alpha \beta} \partial V(r) /\partial \zeta_\beta \, ,
\label{eq:geod_1p}
\end{equation}
where
\begin{equation}
?N^{\alpha}_{}\beta?(\zeta(s),\dot\zeta(s))=\frac{1}{2}\bar\partial_\beta (?\Gamma^\alpha_{}{\gamma \delta}?(\zeta(s),\dot\zeta(s)) \dot\zeta^\gamma \dot\zeta^\delta) 
\label{non-lin-conn}
\end{equation}
is the non-linear Cartan connection (Appendix~\ref{appendixC}) and
$?\Gamma^{\alpha}^{}_{\beta  \gamma}? =\frac{1}{2} {g}^{\alpha \delta}({g}_{\delta  \gamma,\beta}+{g}_{\delta \beta, \gamma}-{g}_{\beta  \gamma,\delta})$ are the generalized connections 
(with ${g}_{\alpha \beta, \gamma}=\partial_{\gamma}{g}_{\alpha \beta}\equiv \partial g_{\alpha \beta}/\partial \zeta_{\gamma}$, and $\dot\zeta(s)=\partial \zeta(s)/\partial s$), 
${g}_{\alpha \beta}$ are the Finsler metric coefficients 
\begin{equation} \label{prop2}
g_{\alpha  \beta}(\zeta(s),\dot\zeta(s))=\frac{1}{2} \bar\partial_{\alpha} \bar\partial_{\beta} L_q^2(\zeta(s),\dot\zeta(s)) \, ,
\end{equation}
with $\bar\partial_{\alpha}=\partial_{\dot\zeta^\alpha}$ and
\begin{equation} \label{prop2b}
L_q(\zeta(s),\dot\zeta(s))=\mathcal{T} (\dot\zeta(s))/\dot{\zeta}^0  - Q(\zeta(s)) \dot{\zeta}^0 \, 
\end{equation}
with $\mathcal{T} (\dot\zeta)=(1/2) \sum_i^N m_i \sum_{k=1}^3(\dot\zeta_i^k)^2$ ($m_i$ is the particle mass) is the (still non-relativistic) kinetic energy.
In Eq.~\eqref{eq:geod_1p}, 
$V(\zeta(s))$ is the classical potential, and $ Q(\zeta(s))$ is the quantum potential, which is nonlocal and depends on all particle positions.

\section{Quantum potential in relativistic field theory} \label{sec_Dirac}

\subsection{Relativistic field theory in a space-time manifold} 
In field theory, the dynamics of spin-1/2 particles is governed by the Dirac equation for the field operator ${\psi}(x)$ 
\begin{equation}
(i \tilde \gamma^{\mu} \, \mathcal{D}_{\mu}  - m) {\psi} (x)=0  \, 
\label{Eq:fieldop}
\end{equation}
derived from the Dirac action functional~\cite{book_birrell_davies,gieres2016}
\begin{equation} \label{eq:appE_action}
\mathcal{A}_D  \left[\psi(x); ?e^\mu_{}a?(x) \right] = i  \int d^4x \,  \sqrt{|g|} \, \bar{\psi}(x) \tilde \gamma^{\mu} \overset{\leftrightarrow}{\mathcal D_{\mu}}  \psi(x) - m \bar\psi(x) \psi(x) \,. 
\end{equation} 
In Eqs.~\eqref{Eq:fieldop} and~\eqref{eq:appE_action}, $m$ is the particle mass, 
 $\mathcal D_{\mu} = ?E^a_{}\mu? \mathcal D_a = ?E^a_{}\mu? (\partial_a + \omega_a)$, 
 $?e^\mu_{}a?(x)$  is the frame vector field (forming a tetrad) and $?E^a_{}\mu?(x)$ its inverse ($?E^a_{}\mu?  ?e^{\nu}_{}a?=\delta^{\nu}_{\mu}$), 
 $\tilde \gamma^{\mu} = ?e^{\mu}_{}a? \gamma^a$, 
 $\omega_b(x)$ is the spin connection (see Appendix~\ref{appendixA})  and 
 $\alpha \overset{\leftrightarrow}{\mathcal D}_{\mu} \beta=  \alpha (\mathcal D_{\mu} \beta) -  (\mathcal D_{\mu} \alpha)\beta $.
 The label $\mu$ refers to the local reference frame (with zero curvature) while $a$ is associated to the general coordinates of the curved manifold.  
 Note that we use the same symbol $\mathcal{D}$ for the covariant derivative in both coordinate systems; the distinction is made clear from the use of the coordinate labels. 
For a short description of the Dirac equation in curved space-time see Appendix~\ref{appendixA}. 

In the configuration space representation of a many-body system, the multi-time wavefunction $\Psi(x_1,x_2,\dots, x_N)$ in the $N$-particle spinor space of dimension $(\mathbb{C}^4)^{\otimes N}$  is governed by the trivial extension of Eq.~\eqref{Eq:fieldop}
\begin{equation}
i \tilde\gamma_{(i)}^{\mu} \, \mathcal{D}_{(i)\mu} \Psi(x_1,x_2,\dots, x_N) - m_i \Psi(x_1,x_2,\dots, x_N) =0 
\label{eq:multi_dirac}
\end{equation}
where $x_i$ is a point in the 4-dimensional space-time, $i=1,\dots,N$; $\mu=0,\dots,3$; $\tilde\gamma_{(i)}^{\mu}= \mathds{1} \otimes \dots \otimes \tilde\gamma^{\mu} \otimes \mathds{1} \otimes \dots \otimes \mathds{1}$ with $\tilde\gamma^{\mu} $ in position $i$, and similarly  for $\mathcal{D}_{(i)\mu}$.

In Eq.~\eqref{eq:multi_dirac},  $\psi(x)$ denotes a Dirac field (annihilation) operator while $\Psi(x_1,x_2,\dots, x_N)$  is the many-electron wavefunction defined by the overlap of the $N$-particle state $|\Psi \rangle$ with the position (bra-) state $ \langle 0| {\psi}(x_1) \dots {\psi}(x_N) |=\langle x_1, \dots, x_N|$, 
\begin{equation}
\Psi(x_1,x_2,\dots, x_N)=\frac{1}{\sqrt{N!}} \langle 0| {\psi}(x_1) \dots {\psi}(x_N) | \Psi \rangle \, ,
\label{Eq:psiwf}
\end{equation}
and therefore
\begin{align}
 |\Psi \rangle &= \int d^4 x_1 \dots d^4 x_N \,   | x_1 \dots x_N \rangle   \langle x_1 \dots x_N |\Psi \rangle \notag \\
   &=  \int d^4 x_1 \dots d^4 x_N \,    \Psi(x_1,x_2,\dots, x_N)  | x_1 \dots x_N \rangle \, .
\end{align}
The anti-commutation relations $\{\psi_{\alpha}(x),\psi_{\beta}(y)\}=\{\psi^{\dagger}_{\alpha}(x),\psi^{\dagger}_{\beta}(y)\}=0$, and $\{\psi_{\alpha}(x),\psi^{\dagger}_{\beta}(y)\}=\delta_{\alpha\beta}\delta^{(3)}(\text{x}-\text{y})$ apply to the Dirac field operators with spinor indices $\alpha$ and $\beta$ ($\text{x}$ and $\text{y}$ are the spatial components of $x$ and $y$, respectively).

So far, the theory is formulated for the case of free, non-interacting, Dirac particles. 
However, we can apply perturbation theory for field operators to describe particle interactions and then use Eq.\eqref{Eq:psiwf} to recover the interacting many-particle wavefunction from the interacting field operators.
For example, the interaction with the electromagnetic field (as discussed below) is obtained with the following substitution 
\begin{equation} \label{eq:covariant_derivative_dirac_interacting}
\mathcal D_{a} \rightarrow \mathcal D_{a}=\partial_{a} + \omega_{a} -i e A_{a} 
\end{equation}
in the global (curved) coordinate system, where $A_a$  denotes the electromagnetic vector potential. 
It is important to stress that all quantum effects are described through the action of the system many-body wavefunction $\Psi(x_1,x_2,\dots, x_N)$ on the geometry of the space-time as discussed in the following sections.

\subsection{Lorentz invariant quantum potential} 

In this work, we propose a GQD theory in which both quantum field theory and general relativity act on the curvature of space-time. 
To this end, as described in refs.~\cite{tavernelli_geom,tavernelli_geom2}, we need to define a quantum potential for the Dirac spin wavefunction in Eq.~\eqref{Eq:psiwf}.
The quantum potential is given by~\cite{Dewdney_1986,Castro_1992,Fanchi_2000} 
\begin{equation}
Q(\bar{x}_N;x_i)=-\frac{\hbar^2}{2m_i} \frac{\partial_{a_i} \partial^{a_i} R(\bar{x}_N)}{R(\bar{x}_N)}
\label{eq_quantum_field_potential}
\end{equation}
where $\bar{x}_N=(x_1,x_2,\dots, x_N)$, $a_i,b_i=0,\dots,3$; $i=1,\dots,N$,
\begin{equation}
R(\bar{x}_N)= (J^{a_1 \dots a_N} (\bar{x}_N) J_{a_1 \dots a_N} (\bar{x}_N) )^{1/2}
\end{equation}
and 
\begin{equation}
J^{a_1 \dots a_N} (\bar{x}_N) = 
\langle \Psi | 
: \frac{1}{N !} 
\bar{\psi}(x_1) \gamma^{a_1} \psi(x_1)
\dots
\bar{\psi}(x_N) \gamma^{a_N} \psi(x_N)
:
| \Psi \rangle \, 
\label{eq:J}
\end{equation}
is the generalized multi-tensor covariant current density (the notation `$: \, \, :$' stands for normal ordering and $\bar{\psi}=\psi^\dagger \gamma^0 $). 
According to~\cite{Dewdney_1986,Hiley_2012}, this would correspond to the `density' component of the quantum potential, which will drive the particle trajectory.
Note that $R^2(\bar{x}_N)$ corresponds to the covariant probability density in the rest frame while the 0-component $J^{0 \dots 0} (\bar{x}_N) J_{0 \dots 0} (\bar{x}_N)$ is not Lorentz invariant~\cite{hollandbook}. 
Since $\partial_{a}=g_{a b} \partial^{b}$, the quantum potential depends on the space-time geometry. 
The amplitude $R(\bar{x}_N)$ in Eq.~\eqref{eq_quantum_field_potential} should not be confused with the scalar curvature of space-time.

\section{Finsler Lagrangians and Finsler metrics for quantum trajectories in general relativity} 
\label{sec_Finsler}

\subsection{Quantum point-particle dynamics and the geometry of space-time}
In Bohmian mechanics, particles follow deterministic paths driven by the combined action of the classical and quantum potentials.
In the quantum field theory extension formulated above, the Bohmian potential is fully determined by the quantum fields, which define the (quantum)  system Lagrangian  that governs the dynamics of the associated  particles. 
Note that in this picture, the fields describe the geometry of space-time and not the particles themselves, which remain point-like in nature and follow deterministic trajectories.
All quantum effects including entanglement and nonlocal correlations are therefore  `confined' to the quantum potential, $Q(\bar{x}_N;x_i)$, acting on a given individual particle $i$. 
This has an important implication on the nature of the geometry of the underlaying space-time.
In quantum theory the symmetrization (or anti-symmetrisation) of the wavefunction for indistinguishable particles hampers the one-to-one assignment of each individual particle to a one-particle subspace 
of the full Hilbert space~\cite{dieks2011}.
In fact, in quantum mechanics all indistinguishable particles have the same one-body density delocalized over the physical space~\cite{note_on_one-body-density}. 
On the other hand, dealing with kinematically independent classical particles enables the decomposition of the many-particle tangent bundle into a direct product of one-particle subspaces: $TM=(TM_i)^N$, where each $M_i$ is the 4-dimensional base manifold associated to particle $i$.
This is clearly possible when the metric tensor of Eq.~\eqref{prop2} has a block diagonal form in the particle subspaces $g_{\alpha \beta}(\bar{x}_N)=\bigoplus_{i=1}^N g_{a_i b_i}(\bar{x}_N)$ with $\alpha,\beta=0,\dots,4N-1$; $a_i,b_i=0,\dots3$; and $i=1,\dots,N$.
The generic single particle subspace manifold referred as $M_0$ equipped with the Finsler function $F_0^{(i)}$ given in Section~\ref{subsect:Finsler_metric} defines the Finsler space ($M_0, F_0^{(i)}$) where the dynamics takes place.
The point-particles are represented as  a set of points in the 4-dimensional \textit{physical} space-time (instead of a single point in the $4N$ dimensional configuration space) and their dynamics follow geodesic curves in the corresponding 8-dimensional Finsler space curved through the action of the quantum potential and of the energy momentum tensor 
(formulations in configuration space are also possible but they will not be discussed in this work).
In the following, $a,b,c \in \{0,1,2,3\}$ label the coordinates $\{x_i^a\}$ of a generic single particle $i$ in $M_0$, $\bar a, \bar b,\bar c \in \{4,5,6,7\}$ the corresponding fiber space $\{ y_i^{\bar a} \}$, and $A,B,C \in \{0,1, \dots, 7\}$. $TM_0$ refers now to the single particle tangent bundle.

The use of the many-time formulation with a time variable $x_i^0$ for each particle introduces the problem of the definition of simultaneity in GR. 
While in special relativity simultaneity acquires a clear physical meaning when an inertial frame is specified, in GR there is no global inertial frame and therefore the concept of `equal-time events' remains ambiguous. 
However, as discussed in ref.~\cite{Misnerbook}, in GR the concept of simultaneity can be replaced by the more general one of `events belonging to the same three-dimensional space-like hypersurface'~\cite{Duerr1999,Duerr2014,Nicolic2012}. 
In this case, events belonging to the same space-like hypersurface occur at the same \textit{global} time, $s$.
The time variable,  $x^0_i(s)$, associated to each $i$-th particle of the system is therefore parametrized by $s$, which measures the progress of the dynamics of all particles.
Further, the families of space-like hypersurfaces corresponding to different global times $s$ define a slicing or foliation $\mathcal{F}$ of space-time with slices (or leafs) $\Sigma_s$.
The ensemble of the coordinates of all particles in the system is therefore characterized by the extended configuration space vector $\bar{x}_N^{\mathcal{F}}=(x^{\Sigma_s}_1,x^{\Sigma_s}_2, \dots, x^{\Sigma_s}_N)$ with $x^{\Sigma_s}_i=(x^0_i(s),x_i^1(s),x_i^2(s),x_i^3(s))$ on the hypersurface $\Sigma_s$ of the foliation $\mathcal{F}$. 
(It is worth mentioning, that - even though rigorous - this definition of  the foliation $\mathcal{F}$ is somehow arbitrary~\cite{Duerr2014,Misnerbook}).

In GR, the geometry of space-time is described by the Lorentzian manifold $({M}_0, g(x))$ of dimension $3+1$. 
At each point $x$ in space-time the metric $g(x)$ defines a linear space ($T_x {M}_0$, $g(x)$) which is locally isometric to the Minkowski space-time $\mathbb{R}^{3+1}$ and defines the \textit{null} cone 
$
\mathcal{N}_x=\{ X \in {T}_x \mathbb{R}^{3+1} , g(x)(X,X) = 0 \}.
$
In particular, the corresponding local Minkowski frame defines a spacelike plane of simultaneity that bisects the cone $\mathcal{N}_x$ and that coincides locally with the selected hypersurface $\Sigma_s$. 
In the following, we will extend the (pseudo-) Riemann geometry on Lorentz space-time to Finsler geometry in the tangent bundle $TM_0$.

\subsection{Finsler function for the non-interacting case}
\label{subsect:Finsler_metric}

For a specific particle of interest we associate a geodesic curve $s \mapsto x_i(s)=\zeta_i(s)$ with components $(\zeta_i^0(s),\dots,\zeta_i^3(s))$ and velocities $s \mapsto y_i(s)=\dot\zeta_i(s)$. As already stated above, the nonlocal quantum potential as well as the system Lagrangian and Finsler functions depend on the position and momenta of all particles considered.

The relativistic covariant point particle Lagrangian~\cite{Feynman1995} corresponding to Eq.~\eqref{prop2b} is (see also~\cite{Kostelecky2011})
\begin{equation} \label{matter_action}
L_{\text{rel}}(\zeta,\dot\zeta)=\frac{1}{2}\sum_i m_i  \int 
 d^4x \, \delta^4(x-\zeta_i(s))  
\sqrt{-g_{ab}(x) \frac{d \zeta_i^a(s)}{ds}\frac{d \zeta_i^b(s)}{ds} } \, ,
\end{equation}
where $s$ is the `global' time defined by the covariant foliation $\mathcal{F}$, $\mathcal{A}=\int ds L$, and $N$ is the number of particles in the system. 
We therefore write the covariant quantum Lagrangian for the particle $i$ as (see also Appendix~\ref{appendixB})
\begin{equation} \label{matter_action2}
L_{0}^{(i)} (\zeta,\dot\zeta)=  \int 
d^4x \, \delta^4(x-\zeta_i(s)) 
\left[ 
\left(
\frac{1}{2}m_i - \frac{Q(\zeta;\zeta_i)}{\sqrt{-g}}
\right)
 \sqrt{-g_{ab}(x) \frac{d \zeta_i^a(s)}{ds}\frac{d \zeta_i^b(s)}{ds}} 
\right] \,.
\end{equation}
In Eq.~\eqref{matter_action2}, $g$ is the determinant of the tensor field $g_{ab}(x)$ (with signature ($-,+,+,+$)) and is determined from variation of the gravitation action in Eq.~\eqref{action}.
Note that the classical limit $(\sqrt{\sum_{k=1}^3(\dot\zeta_i^k)^2} \ll c , \forall i)$ of the contribution involving the quantum potential coincides with the `non-relativistic' term in Eq.~\eqref{prop2b}, since
in this limit (Newton's law with gravitational potential $V^g(x)$ generated by all particles in the system, \cite{Zee_gravitation}) we have $g_{ab}=\text{diag}(g_{00}=(-(1+V^g(\zeta)),1,1,1)$ and therefore $g_{00}=\det(g_{ab})=g$.

In order to make this theory compatible with GR in the classical limit, we define the Finsler function as  
$F^{(i)}_{0} (\zeta,\dot\zeta)=\frac{2}{m_i} L^{(i)}_{0}(\zeta,\dot\zeta)$, 
\begin{equation} \label{Finsler_0i}
F_{0}^{(i)} (\zeta,\dot\zeta)=  \int 
d^4x \, \delta^4(x-\zeta_i(s)) 
\left[ 
\left( 
1-\frac{2 Q(\zeta;\zeta_i)}{m_i \sqrt{-g}}
\right)
 \sqrt{-g_{ab}(x) \frac{d \zeta_i^a(s)}{ds}\frac{d \zeta_i^b(s)}{ds}} 
\right] \,.
\end{equation}

The corresponding Finsler metric is therefore
\begin{align} 
\tilde{g}^0_{ab}(\zeta_i)&=\frac{1}{2} \bar\partial_{a} \bar\partial_{ b} (F_{0}^{(i)}(\zeta,\dot{\zeta}) )^2 \notag \\
                   &=\frac{1}{2} \bar\partial_{a} \bar\partial_{ b} 
                   \left(
                           -g_{ab}(\zeta_i) \, 
                            \left(1 - \frac{2 Q(\zeta;\zeta_i)}{m_i \sqrt{-g}}\right)^2 
                            \frac{d \zeta_i^a(s)}{ds}\frac{d \zeta_i^b(s)}{ds}
                      \right)  \notag \\ 
                      &=  - g_{ab}(\zeta_i)    \left(1 - \frac{2 \, Q(\zeta;\zeta_i)}{m_i \sqrt{-g}}\right)^2   \label{metric_vac}    \, ,
\end{align}
which is equivalent to a modified Lorentzian metric in space-time. 
Note the change of signature, which is consistent with the different definitions of the line elements in GR (Eq.~\eqref{matter_action}) and in Finsler spaces.
$\tilde{g}^0_{ab}$ corresponds to the 4-dimensional subspace associated to the particle $i$ while being a function of the entire $4N$-dimensional base manifold through the quantum potential $Q(\zeta;\zeta_i)$.
The metric in Eq.~\eqref{metric_vac} is smooth and non-degenerate in $TM_0$ (see also~\cite{Pfeifer2013}). 
Since there is no dependence of the metric tensor from the fiber coordinates, $y$, the dynamics takes place in the familiar Lorentzian manifold and the Einstein field equations remain form-invariant
\begin{equation} \label{einstein_field_eq}
?R_{ab}?-\frac{1}{2} R \, \tilde{g}^0_{a  b}= k_{\mathcal{G}} T_{ab}   \, , \\
\end{equation} 
with the modified metric defined in Eq.~\eqref{metric_vac}, which includes quantum effects through the action of the quantum potential.

\subsection{Finsler function for the interacting case}

The interaction of the particles with an external potential can be included with the addition of an interaction term of the form
\begin{equation} \label{cq_action_inter}
F^{(i)}_{\text{int}} (\zeta,\dot\zeta) = e A_a(\zeta_{i}, \dot \zeta_{i}) \dot{\zeta}^a_i
\end{equation}
as for the case of the interaction with an electromagnetic field $A(x,y)$ 
(the generalization of the 4-potential $A(x)$ to the Finsler space, see also Section~\ref{subsection:interaction_field})
For each particle $i$, the full covariant Finsler function takes the form
\begin{equation} \label{cq_action_inter}
F^{(i)}_{\mathscr{I}} (\zeta,\dot\zeta) =  F^{(i)}_{0} (\zeta,\dot\zeta) +  e A_c(\zeta_{i},\dot \zeta_{i}) \dot{\zeta}^c_i.
\end{equation}
Finslers spaces with this functional form are  known as Randers spaces~\cite{Randers1941,Bucataru2007,Pfeifer2013}.
Note that the theory is defined up to the tensor field $g_{ab}(\zeta)$, which describes the effects of the gravitational field and is determined through variation of the action functional derived in the next section.

The covariant relativistic Finsler metric in the $TM_0$ subspace associated to a particle $i$ (with components $a\in \{0,1,2,3\}$) is defined as
\begin{align} \label{cq_gF}
\tilde{g}_{a  b}(\zeta_i,\dot{\zeta_i})=& \frac{1}{2} \bar\partial_{a} \bar\partial_{ b} (F^{(i)}_{\mathscr{I}}(\zeta,\dot{\zeta}))^2 \notag \\
                              =& \frac{1}{2} \bar\partial_{a} \bar\partial_{b}  (\sqrt{\tilde{g}^{0}_{ab} (\zeta) \dot{\zeta}_i^a \dot{\zeta}_i^b} + e A_c(\zeta_i,\dot \zeta_i) \, \dot{\zeta}_i^c )^2 \, ,
\end{align}
where  $F^{(i)}_{\mathscr{I}}(\zeta,\dot{\zeta})$ is the sum of a non-interacting term $F^{(i)}_{0} (\zeta,\dot{\zeta})$ with the interaction $e A_a(\zeta_i,\dot \zeta_i) \dot{\zeta}_i^a$ as described in Eq.~\eqref{cq_action_inter}. 
In this case the null structure 
$N=\{ (\zeta_i,\dot{\zeta}_i) \in {TM}_0 : F^{(i)}_{\mathscr{I}}(\zeta,\dot{\zeta}) =0\}$ 
does not coincide with the \textit{classical} Lorentzian null space for which 
$\tilde{g}^0_{ab} (\zeta) \dot{\zeta}_i^a \dot{\zeta}_i^b=0$. 
In addition, in order to guarantee the positivity of 
$F^{(i)}_{\mathcal{R}}(\zeta,\dot{\zeta})\equiv\sqrt{\tilde{g}^0_{ab} (\zeta) \dot{\zeta}_i^a \dot{\zeta}_i^b} + e A_a(\zeta_i,\dot \zeta_i) \, \dot{\zeta}_i^a$ 
one can restrict the $g$-norm of the one-form $A(x,y)$ so that $\tilde{g}^0_{ab} A^aA^b<1$ on $M$~\cite{Bao_2004}. 
This guarantees that $\tilde{g}_{a  b}(\zeta,\dot{\zeta})$ in Eq.~\eqref{cq_gF} is positive defined at all $y\neq0$.
For more details on the geometrical properties of Finsler-Randers spaces and their extensions to the so-called `Finsler space-time' the reader is referred to the specialized literature~\cite{Bucataru2007,Pfeifer2011}.

\section{Gravitational field equation for deterministic quantum trajectories in Finsler space} \label{sec_action}

The gravitation field equations can be derived from the variation of the generic action  
\begin{equation}
\mathcal{A}(g, \psi, A, \dots)=\int d^4x \, \mathcal{L} (g, \psi, A, \dots)=\mathcal{A}_g(g)+\mathcal{A}_m( \psi, A, \dots)
\end{equation}
where $ \mathcal{L}$ is the Lagrangian density, 
\begin{equation} \label{Einsten_action}
\mathcal{A}_g(g)=-\frac{1}{16 \pi G} \int d^4x \, R \sqrt{-g} \, ,
\end{equation}
$R$ is the Riemannian scalar curvature and  the matter-action $\mathcal{A}_m( \psi, A, \dots)$ is a functional of the matter field ($\psi$), the electromagnetic field ($A$) and all other known interaction fields ($G$ is the universal gravitational constant, and we use the units $\hbar=c=1$).
In classical GR, Einstein's field equations are derived by setting $\delta \mathcal{A}(g, \psi, A, \dots)=0$.
The generalization of this procedure to classical (non-quantum) Finsler spaces has been already successfully explored in the literature~\cite{Pfeifer2013,Minas2019}.

Using the development of the previous sections, we now derive a generalized equation of motion for the dynamics of quantum particles subjected to gravitational forces in the extended Finsler cotangent bundle $T^*M_0$. 
The first assumption we make is that all quantum mechanical effects (in particular entanglement) are captured by the quantum potential and therefore the dynamics of the single particle can be resolved in its own sector of the $N$-particle tensor space defined in Eq.~\eqref{eq:multi_dirac}.

The geodesic equation (Eq.~\eqref{eq:geod_1p}) combined with the Finsler metric in Eq.~\eqref{cq_gF} define the trajectory of a relativistic particle associated to a quantum field, which for instance can be visualized by the trace in a bubble chamber between the instants of creation and annihilation.
Here we use what we believe is the simplest extension of GR to Finsler spaces. However, it is important to stress that there is still an ongoing debate in the literature about which type of Finsler extension of Einstein's field equations are viable under the most general conditions (see for instance~\cite{Javaloyes_2018,Hohmann_2019}).

The extended Einstein's field equations are derived from the variation of the action (see G. Minas and coworkers~\cite{Minas2019}) 
\begin{equation}
\mathcal{A}^F_{g}=\frac{1}{16 \pi G} \sum_{i=1}^N \int ds \, d^8u\, \delta^4(x-\zeta_i(s)) \, \delta^4(y-\dot\zeta_i(s))   \sqrt{\det \mathcal{G}} \, (R-S) \, ,\label{action}
\end{equation}
with $d^8u=dx^0\wedge dx^1 \dots \wedge dy^2 \wedge dy^3$, 
where
$\mathcal G$ is the Finsler-Sasaki metric defined as 
\begin{equation}
\mathcal G=\tilde{g}_{ab}(x,y) dx^a \otimes dx^b + \tilde{v}_{\bar a \bar b}(x,y) \delta y^{\bar a} \otimes \delta y^{\bar b}
\label{eq:Finsler-Sasaki}
\end{equation}
and $R$ and $S$ are the curvature scalars (fully contracted  curvature (1-3)-tensor components) in the horizontal-vertical split (for this particular choice of the linear connection see seminal work of R. Miron~\cite{Miron_1986}  and Appendix~\ref{appendixC}).
Note that, while we are using the same symbol as in Eq.~\eqref{Einsten_action}, the scalar curvature $R$ refers to the cotangent bundle $T^*M_0$ with Finsler-Sasaki metric.
The field equations for the gravitation metric tensor $g$, which appears in the definition of $g^F$ (Eq.~\eqref{metric_vac}) and $\tilde{g}$ (Eq.~\eqref{cq_gF}) become~\cite{Minas2019}  
\begin{align} 
?R^{c}_{acb}?-\frac{1}{2} (\tilde{g}^{ld} ?R^{c}_{lcd}? + \tilde{g}^{ld}  ?S^{c}_{lcd}? ) \tilde{g}_{ab}&= k_{\mathcal{G}} T_{ab}          \label{cq_einstein1} \\
?S^{\bar{c}}_{\bar{a} \bar{c} \bar{b}}?-\frac{1}{2} (\tilde{v}^{\bar{l}\bar{d}} ?R^{\bar{c}}_{\bar{l}\bar{c}\bar{d}}? + \tilde{v}^{\bar{l}\bar{d}}  ?S^{\bar{c}}_{\bar{l}\bar{c}\bar{d}}?) \tilde{v}_{\bar{a} \bar{b}} &= k_{\mathcal{G}} T_{\bar a \bar b}    \label{cq_einstein2} 
\end{align}
where  indices $a,b$ refer to the base manifold, $M_0$, and $\bar{a}, \bar{b} $ to the fiber space, $k_{\mathcal{G}}=8 \pi G / c^4$, 
$?R^{c}_{acb}?$, $?S^{c}_{acb}?$ are the components of the curvature tensor, $\mathbb{R}$
in the tangent space of $TM_0$ in the horizontal-vertical basis (see Appendix~\ref{appendixC}). 
Furthermore, one can associate the mixed terms $T_{\bar a b}$ and $T_{a \bar b}$ to the corresponding Ricci tensor, giving $^1P_{ab}= k_{\mathcal{G}}T_{\bar a b}$
and $^2P_{ab}= -k_{\mathcal{G}} T_{a \bar b}$~\cite{Bucataru2007,Pfeifer2013}.

Considering the minimal possible extension to Einstein's original formulation, we restrict to the case where 
$T_{\bar a b}=T_{a \bar b}=T_{\bar a \bar b}=0$ and the energy-momentum tensor from 
the particle dynamics is obtained through variation of the corresponding action 
\begin{align}
\mathcal{A}^{(p)}_m = \sum_i m_i \int d^4 x \,  d^4 y \, ds \, 
& \delta^4(x-\zeta_i(s)) \,   \delta^4(y-\dot\zeta_i(s))  \, \notag \\
& \times \left(  \sqrt{\tilde{g}_{ab} \frac{d\zeta^a_i}{ds}  \frac{d\zeta^b_i}{ds}}
+ 
  \sqrt{\tilde{v}_{\bar{a}\bar{b}}  \frac{d\dot\zeta^{\bar{a}}_i}{ds}  \frac{d\dot\zeta^{\bar{b}}_i}{ds}}  \right)
\end{align}
with respect to $\tilde{g}_{ab}$ and $\tilde{v}_{\bar{a}\bar{b}}$. 
This gives  (see Appendix~\ref{appendixD}) 
\begin{align}
T^{ab}(x,y)=
\frac{1}{\sqrt{-\tilde g}} \sum_i m_i  \int ds \,  
\delta^4(x-\zeta_i(s)) \, 
\dot\zeta_i^{{a}}(s)
 \dot\zeta_i^{{b}}(s)
\label{eq:Tab}
\end{align} 
and 
\begin{equation}
T^{\bar{a}\bar{b}}(x,y)=\frac{1}{\sqrt{-\tilde v}} \sum_i m_i  \int ds \,  
\delta^4(y-\dot\zeta_i(s)) \, 
\ddot\zeta_i^{\bar{a}}(s)
 \ddot\zeta_i^{\bar{b}}(s) \, .
\label{eq:Tab0}
\end{equation}
The Dirac field contribution to the energy-momentum tensor (to be added to Eq.~\eqref{eq:Tab}) is given by
\begin{equation}
T^{ab} (x) =\langle  \Psi | t^{ab}(x)  | \Psi \rangle \,  ,
\end{equation}
where (see Appendix~\ref{appendixE}) 
\begin{equation} \label{eq:Dirac_e-m_tensor_oper}
t^{ab} (x) =\frac{i}{2} :  \bar{\psi}(x) ( \gamma^a \overset{\leftrightarrow}{\mathcal{D}^b}  +  \gamma^b \overset{\leftrightarrow}{\mathcal{D}^a} ) \psi(x) : 
\end{equation}
is the  energy-momentum operator of the Dirac fields with $\mathcal{D}_a=\partial_a + \omega_a-i e A_{\mu}$ defined in Eq.~\eqref{eq:covariant_derivative_dirac_interacting} and Appendix~\ref{appendixA}.
The time evolution of the each Dirac field component is derived from the Euler-Lagrange equations applied to $\mathcal{L}_D$ (Eq.~\eqref{eq:appE_action}) while the particles simply follow the geodesic paths given by the metric $\mathcal{G}$ in space-time.

Eqs.~\eqref{cq_einstein1}-\eqref{eq:Tab0} together with the definition of the Finsler space function $F^{(i)}_0$ (Eq.~\eqref{Finsler_0i}), the corresponding metric $\tilde{g}$ (Eqs.~\eqref{metric_vac} and~\eqref{cq_gF}) and the quantum potential $Q(\zeta;\zeta_i)$ (Eq.~\eqref{eq_quantum_field_potential}) constitute the main results of this work.

\subsection{The interacting field contribution to the energy momentum tensor and its consequences in cosmology}
\label{subsection:interaction_field}
Even though a thorough investigation of the extension of the energy momentum tensor to the tangent bundle $TM_0$ goes beyond the scope of this work, we can give a flavour of its implications by looking at the extended electromagnetic field theory. 
In the case of charged particles, the electromagnetic interaction is mediated through the action of the 4-vector potential $A(x)$.
In Finsler space, $A(x)$ acquires an extra dependence on the velocity variables where $A(x,y)$ is 0-homogeneous in $y$.
Note that the field $A(x,y)$ depends on the coordinate of the physical space-time tangent bundle (of dimension $4 \times 4$).
For more information on the physics of the extended Maxwell equations for the field $A(x,y)$ we refer to the specialized literature~\cite{Bucataru2007,Voicu2011}.
The corresponding generalized Faraday 2-form is defined as
$F= dA$ or, in local coordinates~\cite{Voicu2011}, 
\begin{equation}
F=\frac{1}{2} F_{ab} \, dx^a \wedge dx^b + F_{a \bar{b}} \, dx^a \wedge \delta y^{\bar{b}}
\end{equation}
with components 
$F_{ab}=\delta_a A_b-\delta_b A_a$ 
and  
$F_{a\bar{b}}=-\bar\partial_{\bar{b}}A_a$ (see Appendix~\ref{appendixC} for the definitions of $\delta_a$  and $\bar\partial_{\bar{a}}$).
The sources of the electromagnetic field are given by the 4-currents~\cite{Weinberg_book} defined on the spacelike hypersurface $\Sigma_s$ in the usual way, 
\begin{equation}
j(\bar{x}^{\Sigma_s}) = \sum_{i}  e \, \delta^3(\bar{x}^{\Sigma_s} - \bar{\zeta}^{\Sigma_s}_i) \, \dot{\zeta}^{\Sigma_s}_i \, 
\label{eq:current}
\end{equation}
where $e$ us the particles charge and $\zeta^{\Sigma_s}_i$ are the 4-coordinates of the test particle $i$ in $\Sigma_s$ at a given value of the global time $s$, which  
 are solutions of the geodesic equation with metric $\mathcal{G}$ (Eq.~\eqref{eq:Finsler-Sasaki}) 
defined by Eqs.~\eqref{cq_einstein1} and \eqref{cq_einstein2} 
(here we use $\bar{x}$ for the spatial part of the 4-component $x$). 
Note that $A(x,y)$ is coupled to the evolution of the particle dynamics through the second term in Eq.~\eqref{cq_action_inter}.
The total current defines the horizontal components $j^{a}(\bar{\zeta}^{\Sigma_s})$ of a vector field in a $TM_0$.
However, the energy momentum conservation law in Finsler geometry implies the appearance of vertical components of the current, which are not present in pseudo-Riemannian GR, and leads to the general form 
$j=j^{a} \delta_a+j^{\bar{a}} \bar\partial_{\bar{a}}$.
For $A_{a}(x)$ function of the $x$-coordinates only, it follows that $j^{\bar{a}}=0$.

The extended energy momentum tensor of the electromagnetic field in $TM_0$ is obtained from the variation of the action~\cite{Voicu2011} ($c=1$)
\begin{equation}
\mathcal{A}^F_{EM}=-\frac{1}{16 \pi} \int \, \sqrt{|\mathcal{G}|} \, F_{AB} F^{AB} \, d^4x \wedge d^4 y \, , 
\label{eq:A_EM}
\end{equation}
with respect to the spacetime metric ($|\mathcal{G}|$ is the determinant of the Finsler-Sasaki metric in Eq.~\eqref{eq:Finsler-Sasaki}). 
This leads to (see Appendix~\ref{appendixF}) 
\begin{equation}
 T=T_{ab}  \, dx^a \otimes d x^b + T_{a\bar{b}}  \, dx^a \otimes \delta y^{\bar{b}} 
\end{equation}
with componets
\begin{equation}
 T_{aA} =\frac{1}{4 \pi} (-?F_A^B? F_{aB} + \frac{1}{4} \tilde{g}_{aA} F_{BC} F^{BC}) \, ,
 \label{eq:TEM}
\end{equation}
where $\tilde{g}_{a\bar{b}}=0$. 
It is interesting to note that while the energy momentum tensor in the base manifold 
\begin{equation}
T'_{a d}=\frac{1}{4 \pi} (-?F_d^b? F_{ab} + \frac{1}{4} \tilde{g}_{ad} F_{bc} F^{bc}) 
\end{equation}
is traceless, $^{(b)}?T'^{a}_{a}?=0$, the additional mixed components 
\begin{equation}
T''_{a d}=\frac{1}{4 \pi} (-?F_d^{\bar{a}}? F_{a\bar{a}} + \frac{1}{4} \tilde{g}_{ad} F_{\bar{a}\bar{b}} F^{\bar{a}\bar{b}}) 
\end{equation}
generate a nonvanishing-trace tensor with a part proportional to $\tilde{g}_{ad}$, 
which is an archetype of a cosmological constant that can be associated to the dark energy~\cite{Labun2010}.

\subsection{The classical limit}
In the classical limit $Q(\bar{x}_N;x_i) \rightarrow 0$ with no interaction ($A_a(x)=0$), the Finsler non-linear connection 
coefficients simplify to $?N^a_b?(x,y)=?\Gamma^a_{bc}? y^c$~\cite{Bucataru2007,Voicu2011}.
Here $?\Gamma^a_{bc} ?$ are the Christoffel symbols and the curvature obtained from $?N^a_b?(x,y)$ becomes the standard Riemann curvature. 
Since we are now confined to the base manifold and its tangent space (the tangent bundle $TM_0$), the mixed Finsler curvature elements in the tangent space of $TM_0$, namely $?S^{c}_{acb}?$ and $?R^{c}_{acb}?$  all vanish.
From Eq.~\eqref{cq_einstein1}, we then recover the classical Einstein's field equations
\begin{equation}
?R_{ab}?-\frac{1}{2} R \, g_{a  b}= k_{\mathcal{G}} T_{ab}   \, . \\
\end{equation} 
The case where gravity can be neglected and the dynamics is controlled by the quantum potential has been already fully discussed in~\cite{tavernelli_geom,tavernelli_geom2}.

\section{Conclusions}
Most approaches to quantum gravity start from Einstein's classical field theory and derive a coherent and, possibly, renormalizable quantization of the gravitational field i.e., the metric tensor in a  pseudo-Riemannian manifold.
In this work, we addressed gravitational quantum dynamics from the opposite perspective, proposing a theory in which all quantum effects are included into the curvature of a non-Riemannian Finsler space. 
Following the work on the geometrization of quantum mechanics presented in refs.~\cite{tavernelli_geom,tavernelli_geom2}, we first generalized the formalism to make the theory relativistic covariant introducing the concept of many-time field theory together with a relativistic invariant definition of `simultaneity' in GR.
We then proceeded with the definition of the quantum mechanical potential that contributes to the metric tensor of the Finsler space. 
The new theory, both in the non-interacting (metric-Finsler) and interacting (Randers-Finsler) frameworks, is fully covariant, incorporates all quantum mechanical effects and reproduces Einstein's classical gravitation theory in the limit in which the quantum potential vanishes.
Finally, the additional components of the energy momentum tensor appearing in Eq.~\eqref{cq_einstein2} allow for the incorporation of new phenomena that still do not have an explanation within classical GR and that are commonly associated with the presence of dark energy.

\begin{appendices}

\renewcommand{\theequation}{\thesection.\arabic{equation}}
\setcounter{equation}{0}
\section{}
\label{appendixC}

For completeness, in this appendix we introduce the main concepts of Finsler geometry. 
A more detailed description of this topic can be found in the literature (see for instance~\cite{Rund59,Bucataru2007,Voicu2011,Pfeifer2013}).
The dynamics takes place in the tangent bundle $TM$ (a special case of a fiber bundle) with base manifold $M$ of dimension $n=4N$ where $N$ is the number of particles in the system, each one described by 4-coordinates $x_i=(x^0_i, x^1_i,x^2_i,x^3_i)$. 
Note that the first coordinate $x^0_i$ (with associated \textit{velocity} $y^0_i$) corresponds the time variable associated to the particle $i$ and parametrized by a global time parameter $s$ (see main text).
The same formalism can be also applied to the single particle formalism with $N=1$ where all many-body effects are included in the potentials and their effects on the geometry of the space-time manifold.  

The non-locality of the quantum potential imposes the extension of the configuration space from the one of a single particle to the full configuration space of all particles considered, which defines a $4N$-dimensional space (the manifold $M$) and the corresponding tangent bundle $TM \equiv P$.
$P$ has basis $(e_1, \dots, e_N, \hat{e}_1, \dots, \hat{e}_N)\equiv (e,\hat{e})$ and corresponding coordinates $(x_1,\dots,x_{{n}}, y_1, \dots, y_{{n}})\equiv ({x},{y})$. 
Finally, the tangent space to the tangent bundle P (T$_u$P) in a point $u \in P$ is associated to the coordinate basis $(\frac{\partial}{\partial x^1}=\partial_1, \dots, \frac{\partial}{\partial x^{n}}=\partial_{n}, \frac{\partial}{\partial y^1}=\bar\partial_1, \dots, \frac{\partial}{\partial y^{n}}=\bar \partial_{n})\equiv(\partial,\bar{\partial})$. 
When dealing with the dynamic of an arbitrary particle, we look at the 4 dimensional sector of the full configuration space ($M$) that deals with that specific particle. For this subspace, we use the same notation as for the full, with the restriction that the coefficients are restricted to the range $a,b,c=0,\dots, 3$.

The Finsler function $F(x,y)$ defines a (0,2)-d metric tensor field
\begin{equation}
g_{ab}(x,y)=\frac{1}{2} \bar\partial_a \bar\partial_b F^2(x,y) \, ,
\end{equation}
and the (0,3)-d Cartan tensor
\begin{equation}
C_{abc}(x,y)=\frac{1}{4} \bar\partial_a \bar\partial_b \bar\partial_c F^2(x,y) \, .
\end{equation}
Note that for $C_{abc}(x,y)=0$ we recover a Riemannian metric space with the metric tensor $g_{ab}(x)$ independent from $y$.
The corresponding non-linear Cartan connection is given by 
\begin{equation}
?N^{a}_{}b?(x,y)=?\Gamma^a_{}{bc}?(x,y) y^c - ?C^a_{}{bc}?(x,y) ?\Gamma^c_{}{pq}?(x,y) y^p y^q
\label{Nab_def}
\end{equation}
(with $?C^a_{}{bc}?(x,y)=g^{ad}(x,y)?C_{}{dbc}?(x,y)$) 
where $g^{ab}(x,y)$ is the inverse of $g_{ab}(x,y)$ and $?\Gamma^a_{}{bc}?= g^{aq}(\partial_b g_{qc} + \partial_c g_{qb} - \partial_q g_{bc}) $ (to simplify the notation we omit the dependence on the coordinates).
The non-linear curvature is then 
\begin{equation}
?R^a_{}bc?=\delta_c ?N^{a}_{}b? - \delta_b ?N^{a}_{}c? \, .
\end{equation}

The connection allows to decompose the tangent space $T_u P$ 
into the vertical space $V_u P$ 
tangent to $T_u M$. 
This induces the transformation $\{\partial_a, \bar\partial_b\} \rightarrow \{\delta_a=\partial_a -?N^{b}_{}a?\bar{\partial}_{b}, \bar\partial_b \}$ in the basis coordinates of $T_u P$. 
At this point, one can define a linear covariant derivative that preserves the horizontal-vertical split of the tangent bundle $P$ without inducing mixing.
In the horizontal-vertical basis, the linear covariant derivative becomes
\begin{align}
\tilde\nabla_{\delta_a} {\delta_b} &= ?{{\tilde\Gamma}}^c_{}{ab}? \delta_c \\
\tilde\nabla_{\delta_a} {\delta_{\bar{b}}} &= ?{{\tilde\Gamma}}^{\bar{c}}_{}{a\bar{b}}? \bar{\partial}_{c} \\
\tilde\nabla_{\delta_{\bar{a}}} {\delta_b} &= ?{{\tilde Z}}^c_{}{\bar{a}b}? \delta_c \\
\tilde\nabla_{\delta_{\bar{a}}} {\delta_{\bar{b}}} &= ?{{\tilde Z}}^{\bar{c}}_{}{\bar{a}\bar{b}}? \bar{\partial}_{c} 
\end{align}
where $a,b,c=0,\dots, 3 $; $\bar{a}, \bar{b}=4,\dots,7$ and 
\begin{align}
?{{\tilde \Gamma}}^c_{}{ab}? &= \frac{1}{2} g^{cq} (\delta_a g_{bq} + \delta_b g_{aq} - \delta_q g_{ab})\\
?{{\tilde Z}}^c_{}{ab}? &= g^{cq} C_{abq} \, .
\end{align}

In the basis $\{\delta_a,\bar\partial_b\}$ the linear curvature (1,3)-tensor on the tangent bundle ($TM$) 
is given by ($\alpha,\beta, \delta,\gamma,\phi=0,\dots, 7)$)
\begin{equation}
?{\mathbb{R}}^{\alpha}_{\beta \gamma \delta}? = 
X_{\delta} ?\Gamma^{\alpha}_{\beta\gamma}? - 
X_{\gamma} ?\Gamma^{\alpha}_{\beta\delta}? +
?\Gamma^{\phi}_{\beta\gamma}? ?\Gamma^{\alpha}_{\phi \delta}? -
?\Gamma^{\phi}_{\beta\delta}? ?\Gamma^{\alpha}_{\phi \gamma}? +
?\Gamma^{\alpha}_{\beta \phi}? ?W^{\phi}_{\gamma \delta}? \, ,
\end{equation}
where
\begin{equation}
X_\alpha = (\delta_a, \bar{\partial}_{\bar a}) \, , \, \, 
?W^{\bar{a}}_{bc}?=?R^{a}_{bc}? \, , \, \, 
?W^{\bar{a}}_{\bar{b}c}?=-\frac{\partial ?N^{a}_{}c?}{\partial y^b} \, , \, \, 
?W^{\bar{a}}_{b\bar{c}}?=\frac{\partial ?N^{a}_{}b?}{\partial y^c} \, 
\end{equation}
and zero otherwise.  
$\mathbb{R}$ decomposes into the following components labelled by different symbols (capital letters) depending on the addressed (horizontal-vertical) sectors~\cite{Bucataru2007}
\begin{align}
& ?{\mathbb{R}}^a_{}{bcd}? = ?{\mathbb{R}}^{\bar a}_{}{\bar{b}cd}? = ?R^a_{}{bcd}? \\
& ?{\mathbb{R}}^a_{}{bc\bar{d}}?  = -?{\mathbb{R}}^a_{}{b\bar{c}d}? = ?{\mathbb{R}}^{\bar{a}}_{}{\bar{b}c\bar{d}}? = - ?{\mathbb{R}}^{\bar{a}}_{}{\bar{b}\bar{c}d}? =  ?P^a_{}{bcd}? \\
& ?{\mathbb{R}}^{a}_{}{b\bar{c} \bar{d}}? = ?{\mathbb{R}}^{\bar{a}}_{}{\bar{b}\bar{c}\bar{d}}? = ?S^a_{}{bcd}?
\end{align}
while the corresponding Ricci tensor components become
\begin{align}
& {\mathbb{R}}_{ab} = R_{ab}, \\ 
& {\mathbb{R}}_{\bar{a}b} = {^1P}_{ab}, \\ 
& {\mathbb{R}}_{a\bar{b}} = - {^2P}_{ab}, \\ 
& {\mathbb{R}}_{\bar{a}\bar{b}} = S_{ab} \, . \\ 
\end{align}

Note that the linear connection is not uniquely defined and therefore alternative definitions can also be formulated~\cite{Pfeifer2013}.
The horizontal part of the curvature $^l\hspace{-0.07cm}R(\delta_a, \delta_b)(.)$ can be easily evaluated 
\begin{align}
^l\hspace{-0.07cm}?R^q_{}{cab}?=\delta_a ?{{\tilde\Gamma}}^q_{}{cb}? - 
\delta_b ?{{\tilde\Gamma}}^q_{}{ca}? + 
?{{\tilde\Gamma}}^q_{}{ma}? ?{{\tilde\Gamma}}^m_{}{cb}? - 
?{{\tilde\Gamma}}^q_{}{mb}?  ?{{\tilde\Gamma}}^m_{}{ca}? 
- ?C^q_{}{cm}? ?R^m_{}{ab}? \, ,
\end{align}
and is related to the original non-linear curvature through the equation 
\begin{equation}
?R^q_{}{ab}?=- \, ^l\hspace{-0.07cm}?R^q_{}{cab}? y^c \, .
\end{equation}
Finally, the corresponding geodesic curve $s \mapsto \zeta(s)$ defined by 
\begin{equation}
\ddot{\zeta}^a + ?N^{a}_{}b?(\zeta,\dot{\zeta}) \dot\zeta^b= 
- {g}^{a b} \partial V(\zeta(s)) /\partial \zeta_b \, ,
\end{equation}
 reproduces  exactly the dynamics in Eq.~\eqref{eq:geod_1p}.

\renewcommand{\theequation}{\thesection.\arabic{equation}}
\setcounter{equation}{0}
\section{}
\label{appendixA}

The generalization of Dirac field equation to curved space-time geometry defined by $g_{ab}(x)$ (in the coordinates $x^a$ of the curved manifold) requires the notion of the so-called tetrad or vierbein formalism~\cite{book_birrell_davies}, which enables 
the definition of a local normal (Minkowskian) coordinates $\tilde{x}^{\mu}$ at a space-time point X. 
In terms of these new coordinates $\tilde x^{\mu}$ the metric tensor simplifies to $\eta_{\mu \nu}$ and 
\begin{equation}
g_{ab}(x)= e^{\mu}_{a}(x)   e^{\nu}_{b}(x) \, \eta_{\mu\nu}
\end{equation}
 where
 \begin{equation}
e^{\mu}_{a}(x)  = \left( \frac{\partial \tilde{x}^{\mu}}{\partial x^{a}}  \right)_{x=X}
\end{equation}
 defines the vierbein $(\mu=0,1,2,3)$.
 
When passing to curve space-time, the covariant derivative of Dirac fields, $\mathcal{D}_a$, acquires therefore an additional term, the so-called spin connection $\omega_a(x)$ defined as~\cite{book_birrell_davies,gieres2016}
\begin{equation}
\omega_a =  \frac{1}{2} \Omega^{\mu \nu} \, e^{b}_{\mu} e_{\nu b,a} \,
\end{equation}
with
$ 
\Omega^{\mu \nu} = \frac{1}{4} [\tilde\gamma^{\mu},\tilde \gamma^{\nu}] \, 
$ 
and 
$ 
\tilde \gamma^{\mu} = ?e^{\mu}_{}a? \gamma^a \, ,
$ 
which leads to (including the electromagnetic coupling term) 
\begin{align}
\mathcal{D}_{\mu} \psi&=  ?E^a_{}\mu? \left( \partial_a \psi+ \omega_{a}  \psi - i e A_a \psi  \right) \\
\mathcal{D}_{\mu} \bar{\psi}&= ?E^a_{}\mu?  \left( \partial_a \psi - \bar{\psi} \omega_{a}   + i e \bar{\psi}A_a \right) \, ,
\end{align}
where $?E^a_{}\mu?(x)$ is the inverse of $?e^\mu_{}a?(x)$ ($?E^a_{}\mu? \, ?e^\mu_{}b? = \delta^a_b$).
The corresponding Dirac field action is given by~\cite{book_birrell_davies}
\begin{equation} 
\mathcal{A}_D  \left[\psi(x), A(x); ?e^\mu_{}a?(x) \right] = i  \int d^4x \, \sqrt{
|g|} \,  \bar{\psi}(x) \tilde \gamma^{\mu} \overset{\leftrightarrow}{\mathcal D_{\mu}}  \psi(x) - m \bar\psi(x) \psi(x) \,,
\end{equation} 
which upon variation with respect to $\bar \psi$ leads to the Dirac equation (see also Eq.~\eqref{Eq:fieldop})
\begin{equation}
(i \tilde\gamma^{\mu} \, \mathcal{D}_{\mu}  - m) {\psi} (x)=0  \, .
\end{equation}

\renewcommand{\theequation}{\thesection.\arabic{equation}}
\setcounter{equation}{0}
\section{}
\label{appendixB}

In this appendix, we describe the process that led to the definition of the covariant quantum Lagrangian defined in Eq.~\eqref{matter_action2}. 
The goal is to find the simplest possible extension of Eq.~\eqref{matter_action}, which has Eq.~\eqref{prop2b} as its non-relativistic limit.  The quantum potential term in Eq.~\eqref{prop2b}, $-Q(\zeta;\zeta_i) \dot{\zeta_i}^0$ (for the particle $i$), can be obtained by adding 
\begin{equation}
L^{0{(i)}}_{Q}(\zeta,\dot \zeta)=-\int d^4x \, \delta^4(x-\zeta_i(s)) 
\frac{Q(\zeta;\zeta_i)}{\sqrt{-g}}
 \sqrt{-g_{ab}(x) \frac{d \zeta_i^a(s)}{ds}\frac{d \zeta_i^b(s)}{ds}} 
 \,.
\end{equation}
to Eq.~\eqref{matter_action}. 
In the Newton limit we can set $\sqrt{\sum_{k=1}^3 (\dot \zeta_i)^2} \ll c $ and therefore 
\begin{equation}
L^{cl (i)}_{Q}(\zeta,\dot \zeta)=
-\int d^4x \, \delta^4(x-\zeta_i(s)) 
\frac{Q(\zeta;\zeta_i)}{\sqrt{-g}} 
 \sqrt{-g_{00}(x)} \dot \zeta_i^0
 \,.
 \end{equation}
 Finally, taking the classical limit of the metric tensor (see \cite{Zee_gravitation}),
  \begin{equation}
g_{ab}(\zeta_i)=
\begin{pmatrix}
-(1+V^g(\zeta_i)) & 0 & 0 & 0 \\
0 & 1& 0& 0 \\
0 & 0 & 1& 0 \\
0 & 0& 0& 1
\end{pmatrix} 
 \end{equation}
for which $g_{00}=\det(g_{ab})\equiv g$ we get the desired non-relativistic (quantum) limit 
 \begin{equation}
L^{cl (i)}_{Q}(\zeta,\dot \zeta) = -Q(\zeta;\zeta_i) \dot{\zeta}_i^0 \, .
 \end{equation}

\renewcommand{\theequation}{\thesection.\arabic{equation}}
\setcounter{equation}{0}
\section{}
\label{appendixD}

The matter field component of the energy-momentum tensor is obtained through the variation of
\begin{equation}
\mathcal{A}^{(p)}_m =\sum_i m_i \int d^4 x \, ds \, 
\delta^4(x-\zeta_i(s))     \sqrt{\tilde{g}_{ab}(x) \frac{d\zeta^a_i}{ds}  \frac{d\zeta^b_i}{ds}}
\end{equation}
that leads to
\begin{align}
T^{ab}&= \frac{2}{\sqrt{|\tilde g|}} \frac{\delta}{\delta \tilde g_{ab}} \mathcal{A}^{(p)}_m \\
& = \frac{2}{\sqrt{| \tilde g|}}  \sum_i m_i \int  ds \, 
 \frac{\delta^4(x-\zeta_i(s)) }{\sqrt{\tilde g_{cd}(\zeta_i)  \frac{d \zeta_i^{c}}{ds}  \frac{d \zeta_i^{d}}{ds}  }}  \frac{d \zeta_i^{a}}{ds}  \frac{d \zeta_i^{b}}{ds} \\
& = \frac{2}{\sqrt{|\tilde g|}} \sum_i m_i \int ds \, 
\delta^4(x-\zeta_i(s))  \frac{d \zeta_i^{a}}{ds}  \frac{d \zeta_i^{b}}{ds}
\end{align}
where the last step is possible by setting $s$ such that $\sqrt{\tilde g_{cd}(\zeta_i)  \frac{d \zeta_i^{c}}{ds}  \frac{d \zeta_i^{d}}{ds}  } =1 $.

\renewcommand{\theequation}{\thesection.\arabic{equation}}
\setcounter{equation}{0}
\section{}
\label{appendixE}

For completeness, in this appendix we summarize the main steps that leads to the energy momentum tensor, $T_{ab}$,  given in Eqs.~\eqref{eq:Tab} and~\eqref{eq:Tab0}.
To this end, we follow the development in Ref.~\cite{Forger2004} (see Theorem 4.2 therein), which starts with the definition of $T_{ab}$ as the variation of the field action with respect to the metric tensor (in a generic $n$ dimensional manifold $M$):
\begin{equation}  
\delta_g  \mathcal{A}  = \delta_g \int d^n x \,  {\mathcal{L}} = -\frac{1}{2} \int d^n x \,  \sqrt{|g|} T^{ab} \delta g_{ab} \, ,
\label{eq:defTab} 
\end{equation}
with $g= \det g$. 
Using the variation identities
\begin{equation}
 \delta g^{ab} = - g^{ac} g^{bd}  \delta g_{cd} 
\end{equation}
for the inverse metric tensor and
\begin{equation}
\delta \sqrt{|g|} =\frac{1}{2} \sqrt{|g|} \, g^{ab} \delta g_{ab}
\end{equation}
for the metric determinant, we obtain
\begin{equation}
\delta_g \int d^n x \, \mathcal{L}  = \int d^n x\,  \left(\frac{1}{2} g^{ab} \mathcal{L} \, \delta g_{ab} + \frac{\partial \mathcal{L}}{\partial g_{ab}} \delta g_{ab} \right)
\end{equation}
from which (comparing with Eq.~\eqref{eq:defTab}) we get
\begin{equation}
 \sqrt{|g|} \, T^{ab} = -2 \frac{\partial \mathcal{L}}{\partial g_{ab}}  - g^{ab}\mathcal{L}
\label{eq:T^ab_of_L}
\end{equation}
or equivalently
\begin{equation}
 \sqrt{|g|} \, T_{ab} = 2 \frac{\partial \mathcal{L}}{\partial g^{ab}}  - g_{ab}\mathcal{L}
\label{eq:T_ab_of_L}
\end{equation}
where we used 
\begin{equation}
 \frac{\partial g^{ab}}{\partial g_{cd}}= -\frac{1}{2} (g^{ac} g^{bd} + g^{ad} g^{bc}) \, .
\end{equation}

For the Dirac field contribution to the energy momentum tensor in a flat space-time geometry, one gets (with $g_{ab}=\tilde g_{ab}$, and $D_a=\partial_a-i e A_a$)
\begin{align}
\mathcal{L}_D &=  \frac{i}{2}  \sqrt{|\tilde g|}  \bar\psi(x) \gamma^b  \overset{\leftrightarrow}{D}_b \psi(x) \notag \\
&= \frac{i}{2}  \sqrt{|\tilde g|} \bar\psi(x) \gamma_a \tilde g^{ab} \overset{\leftrightarrow}{D}_b \psi(x) \notag \\
&= \frac{i}{2}   \sqrt{|\tilde g|} \bar\psi(x) \gamma_a  \tilde g^{ba} \overset{\leftrightarrow}{D}_b \psi(x) 
\end{align}
or equivalently
\begin{equation}
 \mathcal{L}_D =  \frac{i}{2}  \sqrt{|\tilde g|} \bar\psi(x) \gamma^b  \overset{\leftrightarrow}{D}_b \psi(x)  =  
 \frac{i}{2}   \sqrt{|\tilde g|} \bar\psi(x) \gamma_b  \tilde g^{ba} \overset{\leftrightarrow}{D}_a \psi(x) \, ,
\end{equation}
which implies that
\begin{equation}
2\mathcal{L}_D  =  \frac{i}{2}  \sqrt{|\tilde g|}  \left(  \bar\psi(x) \gamma_a  \tilde g^{ba} \overset{\leftrightarrow}{D}_b \psi(x)    + \bar\psi(x) \gamma_b  \tilde g^{ba} \overset{\leftrightarrow}{D}_a \psi(x)     \right)
\end{equation}
and therefore
\begin{equation}
\mathcal{L}_D  =  \frac{i}{4}  \sqrt{|\tilde g|} \left(  \bar\psi(x) \gamma_a  \tilde g^{ab} \overset{\leftrightarrow}{D}_b \psi(x)    + \bar\psi(x) \gamma_b  \tilde g^{ab} \overset{\leftrightarrow}{D}_a \psi(x)     \right) \, .
\end{equation}
Finally, inserting into equation Eq.~\eqref{eq:T_ab_of_L} one gets
\begin{equation}
T_{ab} =  \frac{i}{2}  \left(  \bar\psi(x) \gamma_a   \overset{\leftrightarrow}{D}_b \psi(x)    + \bar\psi(x) \gamma_b  \overset{\leftrightarrow}{D}_a \psi(x)     \right)  -  \frac{\tilde  g_{ab}}{\sqrt{|\tilde g|}} \mathcal{L}_D \, .
\end{equation}

The extension to  curved space-time requires the introduction of the vielbein fields $?e^\mu_{}a?(x)$ (with inverse $?E^{\mu}_{}a?$, such that $?E^{a}_{}\mu? ?e^{\nu}_{}a?=\delta^{\nu}_{\mu}$) related to the metric by the relation $g_{ab}=\eta_{\mu \nu} ?e^{\mu}_{}a? ?e^{\nu}_{}b?$. 
In the following, we only briefly summarize the main results without giving any derivation. For more details see~\cite{book_birrell_davies,gieres2016}.

The variation of the action functional (without mass term) 
\begin{equation} \label{eq:appE_action_D}
\mathcal{A}_D  \left[\psi, A; ?e^\mu_{}a?(x) \right] = i  \int d^4x \, \sqrt{|g|}  \bar{\psi} \tilde \gamma^{\mu} \overset{\leftrightarrow}{\mathcal{D}}_{\mu}  \psi 
\end{equation}
with respect to the frame fields $?E^a_{}{\mu}?$ yields~\cite{gieres2016}
\begin{equation}
T^{ab} (x) =\frac{i}{2}  \left( \bar{\psi} \gamma^a \overset{\leftrightarrow}{\mathcal{D}^b}  \psi + \bar{\psi}  \gamma^b \overset{\leftrightarrow}{\mathcal{D}^a}  \psi \right) \, ,
\end{equation}
which  corresponds to Eq.~\eqref{eq:Dirac_e-m_tensor_oper}, when transformed to the operator space.

\renewcommand{\theequation}{\thesection.\arabic{equation}}
\setcounter{equation}{0}
\section{}
\label{appendixF}

For completeness, we summarize the main steps that lead to the energy momentum tensor for the electromagnetic field in the Finsler space. The derivation follows closely the one of N. Voicu in~\cite{Voicu2011}. 

For the Lagrangian density (with $c=1$) corresponding to Eq.~\eqref{eq:A_EM} 
\begin{equation}
\mathcal{L}= -\frac{1}{16 \pi}  \, \sqrt{\mathcal{G}} \, F_{AB} F^{AB} 
\end{equation}
the partial derivatives
\begin{align}
\frac{\partial \mathcal{L}}{\partial A_{a,b}}&=-\frac{1}{4  \pi} F^{ba} \sqrt{\mathcal{G}} \\
\frac{\partial \mathcal{L}}{\partial A_{a . \bar{b}}}&=-\frac{1}{4  \pi} F^{\bar{b} a} \sqrt{\mathcal{G}}
\end{align}
give
\begin{align}
?{{\tilde{T}}}^b_{}c? &=\frac{1}{4  \pi} (-F^{ba} A_{a,c} + \frac{1}{4} \delta^b_c F_{AB} F^{AB})  \label{eq:AppE_TEM1}\\
?{{\tilde{T}}}^{\bar b}_{}c? &=- \frac{1}{4  \pi} F^{\bar{b} a} A_{a,c} \label{eq:AppE_TEM2}
\end{align}
where we use the notation $X_{a,c} = \frac{\partial X_a}{\partial x^{{c}}}$ and  $X_{a.c} = \frac{\partial X_a}{\partial y^{\bar{c}}}$. 
To symmetrize these expressions, we can add terms derived from divergences of generic tensors without changing the conserved charges~\cite{Voicu2011} .
Using
\begin{align}
\frac{1}{4  \pi} (F^{ba} A_{a,c} + + F^{b \bar{a}} A_{c . \bar{a}}) \sqrt{\mathcal{G}}  &= \frac{1}{4  \pi}  (F^{ba} A_{c} \sqrt{\mathcal{G}})_{,a} + \frac{1}{4  \pi}  (F^{b \bar{a}} A_c \sqrt{\mathcal{G}})_{. \bar{a}} \\
\frac{1}{4  \pi} F^{\bar{b} a} A_{c,a} \sqrt{\mathcal{G}} &=  \frac{1}{4  \pi} (F^{\bar{b}a} A_c \sqrt{\mathcal{G}})_{,a}
\end{align}
in Eq.~\eqref{eq:AppE_TEM1} and Eq.~\eqref{eq:AppE_TEM2}, we finally obtain the energy momentum tensor in Eq.~\eqref{eq:TEM}.
\end{appendices}

 \section*{Conflict of interest}
The author declare no conflict of interest.

\bibliographystyle{spphys}       

\begin{thebibliography}{10}
\bibitem{Quantum_Gravity_Def}
Since the term `Quantum Gravity' is reserved for the theory dealing with the
  canonical quantization of the gravitational field, I will use `Gravitational
  Quantum Dynamics' (abbreviated with GQD) to referred to the geodesic quantum
  trajectory approach described in this work.
\bibitem{note_on_one-body-density}
Without loss of generality, we compute the one-body density corresponding to a
  symmetric (or anti-symmetric) non-relativistic many-body wavefunction
  $\Psi(r_1,\dots,r_N)$ \begin{equation*} \rho(r_i) = \int d^3r_1 \dots
  d^3r_{i-1} d^3r_{i+1} \dots d^3r_N \, \Psi(r_1,\dots,r_N)
  \Psi^*(r_1,\dots,r_N) \, , \end{equation*} which is, by symmetry, the same
  for any choice of $i$ between $1$ and $N$.
\end{thebibliography}

\end{document}